\documentclass[aps,prb,twocolumn,amssymb,amsmath,superscriptaddress]{revtex4-1}
\usepackage{bm}
\usepackage{times}
\usepackage{graphicx}
\usepackage{color}
\usepackage{dcolumn}
\usepackage[colorlinks=true, letterpaper=true, pdfstartview=FitV, linkcolor=blue, citecolor=blue, urlcolor=blue]{hyperref}

\begin{document}
\title{Nonlinear Hall effect induced by internal Coulomb interaction and phase relaxation process in a four-terminal system with time-reversal symmetry}
\author{Miaomiao Wei}
\affiliation{College of Physics and Optoelectronic Engineering, Shenzhen University, Shenzhen 518060, China}
\author{Bin Wang}
\affiliation{College of Physics and Optoelectronic Engineering, Shenzhen University, Shenzhen 518060, China}
\author{Yunjin Yu}
\affiliation{College of Physics and Optoelectronic Engineering, Shenzhen University, Shenzhen 518060, China}
\author{Fuming Xu}
\email[]{xufuming@szu.edu.cn}
\affiliation{College of Physics and Optoelectronic Engineering, Shenzhen University, Shenzhen 518060, China}
\author{Jian Wang}
\email[]{jianwang@hku.hk}
\affiliation{College of Physics and Optoelectronic Engineering, Shenzhen University, Shenzhen 518060, China}
\affiliation{Department of Physics, University of Hong Kong, Pokfulam Road, Hong Kong, China}

\begin{abstract}

We numerically investigate the second-order nonlinear Hall transport properties of a four-terminal system with time-reversal symmetry and broken inversion symmetry. Within the nonequilibrium Green's function formalism, the second-order nonlinear conductances are derived, where the internal Coulomb potential in response to external voltages is explicitly included to guarantee the gauge invariance. For the system with single mirror symmetry $\mathcal{M}_{x}$, nonlinear Hall properties are only observable in the $y$ direction and contributed solely from the second-order nonlinear effect. From the symmetry point of view, the observed nonlinear Hall transport phenomena have one-to-one correspondence with the Berry curvature dipole induced nonlinear Hall effect semiclassically obtained for the same Hamiltonian. In addition to the nonlinear Hall effect originated from symmetries of the system, it is found that the internal Coulomb potential has the same symmetry of the four-terminal system, which gives rise to an extra nonlinear Hall response. Moveover, the phase relaxation mechanism modeled by virtual probes leads to the dephasing-induced nonlinear Hall effect.

\end{abstract}

\maketitle

\section{introduction}

Conventional electric Hall effects in the linear response regime rely on the broken time-reversal symmetry, either by extrinsic magnetic fields or intrinsic magnetic materials. Counter-intuitively, in the nonlinear response regime, nonlinear Hall effect can exist in time-reversal-invariant materials without inversion symmetry, and its origin was attributed to the nonzero Berry curvature dipole (BCD) in the band structure~\cite{L-Fu,Guinea1,NHEREV}. Such BCD-induced nonlinear Hall current is linear in the relaxation time, hence it is an extrinsic Hall effect. Several Weyl semimetal (WSM) materials ~\cite{Yan,Brink} and two-dimensional transition metal dichalcogenides~\cite{Low,Xie2018} were proposed as possible candidates to host Berry curvature dipole or the corresponding nonlinear Hall effect. It was soon confirmed by experiments in monolayer~\cite{Q-Ma} and few-layer WTe$_2$~\cite{K-Kang,S-Xu,Lindenberg}, as well as Weyl semimetal TaIrTe$_4$~\cite{H-Yang}, which are all time-reversal invariant and noncentrosymmetric materials. These experimental progresses have attracted intensive research interests on BCD-related topics~\cite{L-Fu1,Ortix1,HJiang1,HJiang2,Law,Ortix2,Saha,W-Zhang,S-Zhang,Tewari,Ortix3,Guinea}. For instance, either spin-orbit interactions\cite{S-Xu, Q-Ma,K-Kang,L-Fu1,Ortix1,Low,Law} or warping of the Fermi surface\cite{Ortix2} was found necessary to induce nonzero BCD in a time-reversal and inversion-broken system. Alternatively, the merging of a pair of Dirac nodes could also lead to finite BCD in 2-dimensional Dirac semimetals\cite{Saha}. Except the nonlinear electric Hall current, BCD-caused nonlinear transport phenomena include nonlinear thermal effects driven by the temperature gradient~\cite{G-Su,Tewari2} and nonlinear Hall photocurrent in Weyl semimetals~\cite{Yan,W-Zhang}, etc.

While Berry curvature dipole is the band signature in momentum space, the resulting nonlinear Hall signals are detected on multi-terminal planar Hall bars in real space by experiments~\cite{S-Xu,Q-Ma, K-Kang,H-Yang}. Therefore, it is necessary to study the nonlinear transport properties of four-terminal systems with symmetries allowing nonzero Berry curvature dipole. Similar to the disorder effect, phase relaxation processes such as phonon and electron-electron interaction widely exist and have important influence in quantum transport. The virtual probe technique~\cite{but-virtual} is commonly used in modeling the phase relaxation process\cite{datta,but-pump,Xing,Sun}. For example, it was both theoretically predicted~\cite{but-pos} and experimentally verified~\cite{exp-pos} that, the electron dephasing due to virtual probes can lead to a positive cross correlation. In view of the disorder-induced nonlinear Hall effect~\cite{disorderNHE,QTNHE}, it is interesting to evaluate the influence of the dephasing mechanism on the nonlinear Hall effect, which is absent so far.

In this work, we study the nonlinear Hall effect in a time-reversal-invariant four-terminal device with broken inversion symmetry. The second-order nonlinear Hall resistance and Hall current as well as the transverse heat current are numerically investigated within the gauge invariant theory expressed in nonequilibrium Green's function. Since the internal Coulomb potential induced by electric response to external voltages is essential to guarantee the gauge invariance for nonlinear transport~\cite{but1,but22}, it is explicitly included in calculating the second-order nonlinear Hall properties. We find that the induced internal Coulomb potential has the same symmetry of the underlying system. Similar to the semiclassical BCD-induced nonlinear Hall effect, we consider the same 2D massive Dirac Hamiltonian and find one-to-one corresponding nonlinear Hall transport phenomena in the quantum regime, from the symmetry point of view. Specifically, the internal Coulomb potential is discovered to have the same spatial symmetry of the four-terminal system. As a result, an additional nonlinear Hall effect is generated due to this internal Coulomb potential. Because the internal Coulomb potential is always against the external electric field, this additional nonlinear Hall effect reduces the overall nonlinear Hall signal. The dephasing mechanism is also evaluated by employing the virtual probe. Similar to the internal Coulomb potential, the voltage profile of virtual probes also has the same symmetry as the system, which introduces the dephasing-induced nonlinear Hall effect.

The rest of the paper is organized as follows. In section II, the gauge invariant theory is briefly reviewed and the second-order nonlinear conductance is defined. Section III starts with general analysis on relations between the linear and nonlinear conductances of a time-reversal-invariant four-terminal system for two different spatial symmetries. Then the nonlinear Hall resistance and Hall current are numerically calculated along with detailed discussion. In addition, the dephasing effect and temperature influence are evaluated. A summary is finally given in section IV.

\section{Gauge invariant theory and the second-order nonlinear conductance}

For a multi-terminal system, the current in terminal $\alpha$ is calculated from the Landauer-B\"uttiker formula ($\hbar=1$)\cite{but-phys}
\begin{equation}
I_{\alpha} = -q \sum_{\beta} \int_E Tr [\Gamma_{\alpha} G^r \Gamma_{\beta} G^a] (f_\alpha-f_{\beta}),
\label{X2final1}
\end{equation}
where $\int_E \equiv \int (dE/2\pi)$, $\Gamma_\alpha= \Gamma_{\alpha} (E-qV_\alpha)$ is the linewidth function, and $f_\alpha = f_\alpha(E-qV_\alpha)$ is the Fermi distribution function of terminal ${\alpha}$. The retarded Green's function $G^{r} = G^{r}(E,U)$ depends on $U$, an internal self-consistent Coulomb potential that must be included into $G^{r}$ to satisfy the gauge invariant condition. $G^{a} = {G^{r}}^\dag$ is the advanced Green's function. In the Hartree approximation, the retarded Green's function in real space is given by
\begin{equation}
G^r(E,U) = \frac{1}{E-H-qU-\Sigma^r},
\label{X2gr10}
\end{equation}
where $\Sigma^r = \sum_{\alpha} \Sigma^r_{\alpha}(E-qV_{\alpha})$ is the self-energy which depends explicitly on external voltages, and  $\Gamma_\alpha=-2{\rm Im}\Sigma^r_\alpha$. The Coulomb potential $U(x)$ satisfies the following self-consistent Poisson equation,
\begin{equation}
\nabla^2 U(x) = 4\pi i q\int_E [G^<(E,U)]_{xx},
\label{X2poisson0}
\end{equation}
where $x$ labels the position. The lesser Green's function $G^<$ is given by $G^< = G^r \Sigma^< G^a$ with
\begin{equation}
\Sigma^< = \sum_{\beta} i \Gamma_{\beta}(E-qV_{\beta}) f_\beta(E-qV_{\beta}).
\label{X2lesser1}
\end{equation}
Eq.(\ref{X2poisson0}) is, in general, a {\it nonlinear} equation since $G^{r,a}$ depends on $U(x)$. Eqs.(\ref{X2final1}), (\ref{X2gr10}), (\ref{X2poisson0}), and (\ref{X2lesser1}) form basic equations of the general gauge invariant DC transport theory. In quantum transport, one must calculate the Green's function along with the Poisson equation self-consistently. Clearly, the current expressed in Eq. (\ref{X2final1}) is gauge invariant: shifting the potential everywhere by a constant $V$, $U\rightarrow U+V$ and $V_{\alpha}\rightarrow V_{\alpha}+V$, $I_\alpha$ from Eq. (\ref{X2final1}) remains the same if we change the variable $E$ to $E-qV$.

In the weakly nonlinear regime, we expand the Coulomb potential $U(x)$ in the following form,
\begin{equation}
U(x)= U_{eq}(x) + \sum_{\alpha} u_{\alpha}(x) V_{\alpha} + ...  \label{X2char}
\end{equation}
where $U_{eq}$ is the equilibrium potential when there is no external bias, and $u_\alpha(x)$ is the characteristic potential\cite{but1,ma1} which describes the first-order internal response due to the Coulomb interaction to the external bias. As electrons are injected into the system, a nonequilibrium charge distribution is formed due to the long-range Coulomb interaction, which induces the internal Coulomb potential. This induced Coulomb potential maintains the gauge invariance for nonlinear quantum transport~\cite{but22}, and closely related to the characteristic potential $u_\alpha$. Expanding $G^<$ from Eq. (\ref{X2poisson0}) in power of $V_\alpha$, we can derive the equations for all the characteristic potentials. Defining $G^r_0=1/(E-H-qU_{eq}-\Sigma^r_{eq})$ and $\Sigma^r_{eq} = \sum_\alpha \Sigma^r_\alpha(E)$ and using the Dyson equation, we have
\begin{eqnarray}
G^r = G^r_0 + G^r_0 \left[ qU-qU_{eq}
+ \Sigma^r - \Sigma^r_{eq}\right] G^r_0 +\cdots
\nonumber
\end{eqnarray}
with $G^r_0$ the equilibrium retarded Green's function, {\it i.e.}, when
$U=U_{eq}$.
At the lowest order, we obtain from Eq.(\ref{X2poisson0})\cite{but1,wbg2,zhanglei}
\begin{equation}
-\nabla^2 u_\alpha = 4\pi q^2 \frac{dn_\alpha}{dE} - 4\pi q^2 \frac{dn}{dE} u_\alpha, \label{X2u1}
\end{equation}
where $dn_\alpha(x)/dE$ is the {\it injectivity} of terminal $\alpha$\cite{but1,X2foot11,ma},
\begin{eqnarray}
\frac{dn_{\alpha}(x)}{dE} = -\int_E \partial_E f [G^r_0 \Gamma_{\alpha} G^a_0]_{xx}, \label{X2inj}
\end{eqnarray}
and $dn_\alpha(x)/dE$ satisfies the general relation\cite{but22,wbg2,levinson,but1}:
\begin{equation}
\sum_\alpha dn_{\alpha}(x)/dE = dn(x)/dE, \label{X2inj1}
\end{equation}
with $dn(x)/dE$ the local charge density.

Clearly, the first term on the right-hand-side of Eq.(\ref{X2u1}) corresponds to the charge density due to external injection, while the second term describes the induced charge density in the system. Due to the gauge invariance, the characteristic potential $u_\alpha$ follows the sum rule $\sum_{\alpha} u_{\alpha} = 1$\cite{but1}. The Thomas-Fermi approximation\cite{but1} is adopted in deriving Eq.(\ref{X2u1}).

The second-order nonlinear conductance $G_{\alpha \beta \gamma}$ is defined by expanding the current in terms of external bias voltages to the second order\cite{wbg2,X2foot20},
\begin{equation}
I_{\alpha} = \sum_{\beta} G_{\alpha \beta} V_{\beta} + \sum_{\beta
\gamma} G_{\alpha \beta \gamma} V_{\beta} V_{\gamma} + ... \label{X2current}
\end{equation}
with
\begin{eqnarray}
G_{\alpha \beta \gamma} = G^e_{\alpha \beta \gamma}+G^i_{\alpha \beta \gamma}. \label{eq14}
\end{eqnarray}
$G^e_{\alpha \beta \gamma}$ is the external contribution from injected electrons,
\begin{eqnarray}
G^e_{\alpha \beta \gamma} &=& -(q^3/2) \int_E \partial_E f  \delta_{\beta \gamma} {\rm Tr}[(\Gamma \delta_{\alpha \gamma} - \Gamma_\gamma)
\nonumber \\
&\times& (G^a_0 \Gamma_\alpha G^r_0 G^r_0 + G^a_0 G^a_0 \Gamma_\alpha G^r_0 )], \label{X2g111n}
\end{eqnarray}
while $G^i_{\alpha \beta \gamma}$ is the internal contribution from the Coulomb potential,
\begin{eqnarray}
G^i_{\alpha \beta \gamma} &=& q^3 \int_E {\rm Tr}[( G^a_0 \Gamma_\alpha G^r_0 u_\beta G^r_0+
G^a_0 u_\beta G^a_0 \Gamma_\alpha G^r_0) \nonumber \\
&\times& (\Gamma \delta_{\alpha \gamma} - \Gamma_\gamma)] \partial_E f, \label{X2g111n1}
\end{eqnarray}
where $u_\beta$ is obtained from Eq.(\ref{X2u1}).

Finally, we symmetrize the definition of the second-order nonlinear conductance $G_{\alpha\beta\gamma} := (1/2)(G_{\alpha\beta\gamma}+ G_{\alpha\gamma \beta})$. It is easy to see that the gauge invariance is equivalent to $\sum_{\beta} G_{\alpha \beta \gamma} = 0$\cite{but5}. The second-order nonlinear conductance of two-terminal systems has been numerically investigated using the scattering matrix approach\cite{wang11, sheng}.

\section{NUMERICAL RESULTS AND DISCUSSION}

To study the nonlinear Hall effect, we choose the Hamiltonian with time-reversal (TR) symmetry and broken inversion symmetry\cite{L-Fu2},
\begin{equation}
H\left( \mathbf{k} \right)=A{{k}^{2}}+\left( B{{k}^{2}}+\delta  \right){{\sigma }_{z}}+{{v}_{y}}{{k}_{y}}{{\sigma }_{y}}+D{{\sigma }_{x}}, \label{ham}
\end{equation}
where $A$, $B$, and $D$ are system parameters, and ${{\sigma }_{x,y,z}}$ are Pauli matrices. Such a low-energy Hamiltonian describes massive Dirac cones tilted by the $v_y k_y {\sigma }_{y}$ term, which has been discussed in various Dirac systems~\cite{L-Fu,disorderNHE,tiltexpe}. To observe the nonlinear Hall effect, both nonzero $D$ and $v_y$ are required to break the inversion symmetry. Note that this Hamiltonian breaks only the mirror symmetry ${{\mathcal{M}}_{y}}$: $H\left( {{k}_{x}}, {{k}_{y}} \right)\ne H\left( {{k}_{x}},{-{k}_{y}} \right)$ while preserves the mirror symmetry ${{\mathcal{M}}_{x}}$: $H\left( {{k}_{x}},{{k}_{y}} \right)=H\left( -{{k}_{x}},{{k}_{y}} \right)$. The same Hamiltonian has been adopted to investigate the nonlinear Hall effect with the semiclassical Boltzmann approach, where Berry curvature dipole related to the mirror symmetry plays the crucial role. In the tight-binding presentation, this Hamiltonian is expressed in a square lattice as
\begin{eqnarray}
H = \sum\limits_\mathbf{i} {\left[ {\psi _\mathbf{i}^\dag {T_0}{\psi _\mathbf{i}} + \left( {\psi _\mathbf{i}^\dag {T_x}{\psi _{\mathbf{i} + {\mathbf{a}_x}}} + \psi _\mathbf{i}^\dag {T_y}{\psi _{\mathbf{i} + {\mathbf{a}_y}}}} \right) + H.c.} \right]}, \nonumber
\end{eqnarray}
where
\begin{eqnarray}
{{T}_{0}}& = &  -4 T_x +\delta {{\sigma }_{z}}+D{{\sigma }_{x}}, \nonumber \\
{T_x}    & = &  - ( {AI + B{\sigma _z}} )/a^2, \nonumber \\
{T_y}    & = &  T_x - iv_y{\sigma _y}/(2a). \nonumber
\end{eqnarray}
Here $\psi _\mathbf{i}^\dag$ is the creation operator at site $\mathbf{i}$ with $\mathbf{i}=\left( {{\mathbf{i}_x},{\mathbf{i}_y}} \right)$ labeling the lattice site. ${\mathbf{a}_x} = \left( {a,0} \right)$ (${\mathbf{a}_y} = \left( {0,a} \right)$) is the unit vector in the $x$ ($y$) direction with $a$ the lattice constant. In the calculation, we set $A = 0, B = 1, \delta = -0.25, {\upsilon _y} = 1.0, D = 0.1$, and $a=1$.

\begin{figure}[tbp]
\includegraphics[width=8.5cm, clip=]{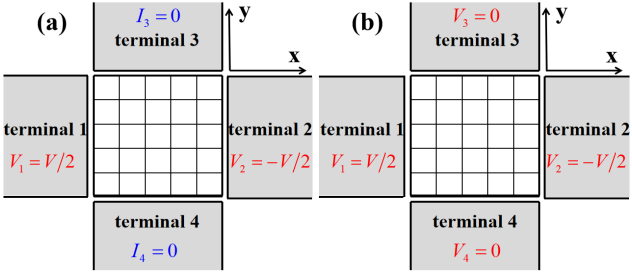}
\caption{(Color online) Schematic diagrams of the 2-dimensional four-terminal system with open (a) or closed (b) boundary condition.}\label{device}
\end{figure}

As depicted in Fig.\ref{device}, two kinds of boundary conditions are considered to evaluate the linear and nonlinear Hall effects in typical four-terminal systems. In Fig.\ref{device}(a), bias voltages are applied in terminals 1 and 2 as $V/2$ and $-V/2$, respectively. By setting $I_3 = I_4=0$, the transverse Hall voltage $V_H$ across terminals 3 and 4 is measured: $V_H = V_3 - V_4$, and the corresponding Hall resistance $R_H = V_H / I_1$ is obtained. We refer this setup as the open boundary condition~\cite{boundcond1,boundcond2}. In Fig.\ref{device}(b), when applying bias voltages in terminals 1 and 2 as $V/2$ and $-V/2$ and maintaining $V_3=V_4=0$, the Hall current $I_H = I_3 - I_4 $ is probed. This case is referred as the closed boundary condition. Similarly, one can also apply a bias across terminals 3 and 4 to measure the Hall resistance between terminals 1 and 2. As will be shown below, both the Hall resistance and Hall current reveal nonlinear Hall characteristics.

\subsection{General discussion}

In this subsection, we will derive and discuss relations among the linear and second-order nonlinear conductances defined in Eq.(\ref{X2current}), when the time-reversal-invariant four-terminal system has different spatial symmetries. By engineering the Hamiltonian expressed in Eq.(\ref{ham}), two spatial symmetries: the inversion symmetry, and the mirror symmetry ${{\mathcal{M}}_{x}}$, are presented in the following cases.

\renewcommand\arraystretch{1.5}
\begin{table}[tbp]
	\begin{center}
		\caption{\label{Symmetric relation I-linear} Relations of the linear conductances in the system with TR and different spatial symmetries. }
		\begin{tabular}{p{3.5cm}<{\centering}|p{5cm}<{\centering}}
\hline\hline
System symmetry     & Conductance symmetry \\ \hline			
TR $\&$ inversion    & ${{G}_{11}}={{G}_{22}}, {{G}_{33}}={{G}_{44}}, {{G}_{13}}={{G}_{14}}={{G}_{23}}={{G}_{24}}$ \\ \hline
TR $\&$ $M_x$ in real space   & ${{G}_{11}}={{G}_{22}},{{G}_{13}}={{G}_{23}}, {{G}_{14}}={{G}_{24}}$\\ \hline\hline			
		\end{tabular}
	\end{center}
\end{table}

\smallskip

\noindent{Case 1. The system with TR and inversion symmetries}

We first study the case where both TR and inversion symmetries are preserved, which corresponds to that the linear term of $k_y$ is dropped in Eq.(\ref{ham}). After the mirror symmetry ${{\mathcal{M}}_{x}}$ transformation$: H\left( {{k_x},{k_y}} \right) \to H\left( {-{k_x},{k_y}} \right)$, external voltages change from $\left( {{V_1},{V_2},{V_3},{V_4}} \right)$ to $\left( {{V_2},{V_1},{V_3},{V_4}} \right)$ for the four-terminal system in Fig.\ref{device}. Subsequently, the currents vary from $I_\alpha$ to $I'_\alpha$ and follow these relations,
\begin{eqnarray}
{I_1} = {I_2}', {I_2} = {I_1}', {I_3} = {I_3}', {I_4} = {I_4}', \label{current_Mx} \\
I_1(V_1,V_2,V_3,V_4) = I_2(V_2,V_1,V_3,V_4). \label{mx}
\end{eqnarray}
Similarly, for the mirror symmetry ${{\mathcal{M}}_{y}}$ transformation $: H\left( {{k_x},{k_y}} \right) \to H\left( { {k_x},-{k_y}} \right)$, bias voltages change from $\left( {{V_1},{V_2},{V_3},{V_4}} \right)$ to $\left( {{V_1},{V_2},{V_4},{V_3}} \right)$ in the device, and the currents satisfy
\begin{eqnarray}
{I_1} = {I_1}', {I_2} = {I_2}', {I_3} = {I_4}', {I_4} = {I_3}', \label{current_My} \\
I_3(V_1,V_2,V_3,V_4) = I_4(V_1,V_2,V_4,V_3). \label{my}
\end{eqnarray}
For a system with inversion symmetry, its transport properties remain invariant under the change from $\left( {{V_1},{V_2},{V_3},{V_4}} \right)$ to $\left( {{V_2},{V_1},{V_4},{V_3}} \right)$, so that the currents obey
\begin{equation}
{I_1} = {I_2}', {I_2} = {I_1}', {I_3} = {I_4}', {I_4} = {I_3}'. \label{current_Mxy}
\end{equation}
From Eqs. (\ref{current_Mx})-(\ref{current_Mxy}) and ${G_{\alpha \beta \gamma }} = {G_{\alpha \gamma \beta }}$, relations among the linear conductances and second-order nonlinear conductances for the system with inversion symmetry are obtained and listed in the first row of Table \ref{Symmetric relation I-linear} and Table \ref{Symmetric relation II}, respectively. Note that there are only seven independent coefficients in the second-order nonlinear conductance in this case.

Now we examine the nonlinear Hall effect up to the second-order in voltage for the system with TR and inversion symmetries, where the Berry curvature is zero. To measure the Hall resistance, the open boundary condition is chosen as in Fig.\ref{device}(a). Using Eqs.(\ref{my}) and (\ref{X2current}), it is easy to show that $V_3=V_4$ is the only physical solution, and hence both the linear and nonlinear Hall resistances are zero along the $y$ direction. This is expected since zero Berry curvature leads to no Hall effect~\cite{berryphase}. Similarly, there is neither no Hall effect along the $x$ direction. If we apply bias voltages in terminals 3 and 4 as $V/2$ and $-V/2$, and maintain zero bias in terminals 1 and 2, it is clear from Eq.(\ref{mx}) that $I_1 = I_2$ indicating zero Hall current. Besides, one can always increase voltages of terminals 3 and 4 by a constant amount $V_3=V_4=V_0$ to make $I_1=I_2=0$, which leads to zero Hall voltage in the $x$ direction. Therefore, for a four-terminal system with TR and inversion symmetries, both the linear and nonlinear Hall signals are zero.

\smallskip

\renewcommand\arraystretch{1.5}
\begin{table}[tbp]
	\begin{center}
		\caption{\label{Symmetric relation II} Relations among the second-order nonlinear conductances in the system with both TR and inversion symmetries. }
		\begin{tabular}{p{8.5cm}<{\centering}<{\centering}}
\hline\hline			
${{G}_{111}}={{G}_{222}},\; {{G}_{333}}={{G}_{444}},\; {{G}_{122}}={{G}_{211}},\; {{G}_{344}}={{G}_{433}}$\\ \hline
${{G}_{112}}={{G}_{212}},\; {{G}_{334}}={{G}_{434}},\; {{G}_{134}}={{G}_{234}},\; {{G}_{312}}={{G}_{412}}$ \\ \hline ${{G}_{113}}={{G}_{223}}={{G}_{114}}={{G}_{224}},\; {{G}_{313}}={{G}_{323}}={{G}_{414}}={{G}_{424}}$\\ \hline ${{G}_{123}}={{G}_{213}}={{G}_{124}}={{G}_{214}},\; {{G}_{314}}={{G}_{324}}={{G}_{413}}={{G}_{423}}$ \\ \hline ${{G}_{133}}={{G}_{233}}={{G}_{144}}={{G}_{244}},\; {{G}_{311}}={{G}_{322}}={{G}_{411}}={{G}_{422}}$\\
\hline\hline	
		\end{tabular}
	\end{center}
\end{table}

\noindent{Case 2. The system with TR and mirror symmetry ${{\mathcal{M}}_{x}}$}

In this case, the mirror symmetry ${{\mathcal{M}}_{y}}$ is broken in real space by adding a potential with symmetry $V(x,y) = V(-x,y)$, and only the mirror symmetry ${{\mathcal{M}}_{x}}$ is preserved. From Eq. (\ref{current_Mx}) and ${G_{\alpha \beta \gamma }} = {G_{\alpha \gamma \beta }}$, we show the relations for the linear and second-order nonlinear conductances in the second row of Table \ref{Symmetric relation I-linear} and Table \ref{Symmetric relation I}, respectively. In this case, only Eq.(\ref{mx}) holds and hence there is no Hall effect along the $x$ direction when bias voltages are applied in $y$ direction. The argument is similar as in Case 1. For the linear Hall effect along the $y$ direction when $V_1$ and $V_2$ are applied, we solve for $V_3$ and $V_4$ under the open boundary condition shown in Fig.\ref{device}(a). From the second row of Table \ref{Symmetric relation I-linear} and the Landauer-B\"{u}ttiker formula, we have
\begin{eqnarray}
I_3 &= G_{31} V_1 + G_{32} V_2 + G_{33} V_3 +G_{34} V_4 =0, \nonumber \\
I_4 &= G_{41} V_1 + G_{42} V_2 + G_{34} V_3 +G_{44} V_4 =0, \nonumber
\end{eqnarray}
from which we find $V_3=V_4=0$ and the linear Hall resistance is zero. If we apply voltages in each terminal as $\mathcal{V}=(V/2,-V/2,0,0)$, i.e., the closed boundary condition, it is easy to show that $I_3=I_4=0$ and hence no linear Hall current as well. We arrive at the same conclusion for linear and second-order nonlinear Hall properties when the mirror symmetry ${{\mathcal{M}}_{y}}$ is broken in momentum space, which is exactly described by Eq.(\ref{ham}).

\renewcommand\arraystretch{1.5}
\begin{table}[tbp]
	\begin{center}
		\caption{\label{Symmetric relation I} Relations between the second-order conductances in the system with broken inversion symmetry.}
		\begin{tabular}{p{8.5cm}<{\centering}}
\hline\hline			
${{G}_{111}}={{G}_{222}},\; {{G}_{122}}={{G}_{211}}$,\; ${{G}_{311}}={{G}_{322}},\; {{G}_{411}}={{G}_{422}}$\\	\hline
${{G}_{112}}={{G}_{212}},\; {{G}_{134}}={{G}_{234}}$,\; ${{G}_{313}}={{G}_{323}},\; {{G}_{413}}={{G}_{423}}$\\ \hline
${{G}_{133}}={{G}_{233}},\; {{G}_{144}}={{G}_{244}}$,\; ${{G}_{114}}={{G}_{224}},\; {{G}_{314}}={{G}_{324}}$\\ \hline
${{G}_{124}}={{G}_{214}},\; {{G}_{414}}={{G}_{424}}$,\; ${{G}_{113}}={{G}_{223}},\; {{G}_{123}}={{G}_{213}}$ \\ \hline\hline	
		\end{tabular}
	\end{center}
\end{table}

\bigskip

Now we discuss the procedure of calculating the second-order nonlinear Hall resistance and current for the system governed by Eq.(\ref{ham}) with the second-order nonlinear theory derived in section II. Both open and closed boundary conditions are considered. When bias voltages are applied in terminals 1 and 2, the open boundary condition corresponds to ${I_3} = {I_4} = 0$, where ${I_3}$ and ${I_4}$ depend quadratically on the external bias as
\begin{eqnarray}
{I}_{3}&=\sum\limits_{\beta}{{{G}_{3\beta}}}{{V}_{\beta}}+\sum\limits_{\beta \gamma}{{{G}_{3\beta \gamma}}}{V}_{\beta }{V}_{\gamma}, \label{currentI3}\\
{I}_{4}&=\sum\limits_{\beta}{{{G}_{4\beta}}}{{V}_{\beta}}+\sum\limits_{\beta \gamma}{{{G}_{4\beta \gamma}}}{V}_{\beta }{V}_{\gamma}. \label{currentI4}
\end{eqnarray}
Four solutions of $V_3$ and $V_4$ are obtained by solving two equations ${I_3} = {I_4} = 0$. Enforcing the following two constraints, only one physical solution is selected: (i) both $V_3$ and $V_4$ are real; (ii) the first-order current is much larger than the second-order one so that higher order terms can be neglected.

Recalling that ${{G}_{31}}={{G}_{32}}$, ${G_{\alpha \beta \gamma }} = {G_{\alpha \gamma \beta }}$ and ${{V}_{1}}=-{{V}_{2}}={V}/{2}\;$, the quadratic equations ${I_3}={I_4}=0$ is simplified as
\begin{equation}
\begin{split}
{{I}_{3}} &= {{G}_{33}}{{V}_{3}}+{{G}_{34}}{{V}_{4}}+\left( {{G}_{311}}+{{G}_{322}}-2{{G}_{312}} \right)V_{1}^{2}\\
&+{{G}_{333}}V_{3}^{2}+2{{G}_{334}}{{V}_{3}}{{V}_{4}}+{{G}_{344}}V_{4}^{2}=0,\\
{{I}_{4}} &= {{G}_{43}}{{V}_{3}}+{{G}_{44}}{{V}_{4}}+\left( {{G}_{411}}-2{{G}_{412}}+{{G}_{422}} \right)V_{1}^{2}\\
&+{{G}_{433}}V_{3}^{2}+2{{G}_{434}}{{V}_{3}}{{V}_{4}}+{{G}_{444}}V_{4}^{2}=0.
\end{split} \label{H-R}
\end{equation}
If we restore the ${\mathcal{M}}_{y}$ symmetry, the inversion symmetry is recovered and the Berry curvature vanishes. As a result, the physical solution is ${V_3}={V_4}$. However, for the system with broken ${{\mathcal{M}}_{y}}$ symmetry, the physical solution ${V_3}$ is not equal to ${V_4}$. Subsequently, the Hall voltage $V_H=V_3-V_4 \ne 0$ and the Hall resistance ${{R}_{H}} ={{{V}_{H}}}/{{{I}_{1}}} \ne 0$.

As for the closed boundary condition, currents $I_3$ and $I_4$ are directly calculated from Eqs.(\ref{currentI3}) and (\ref{currentI4}). Apparently, the linear terms vanish, $I_{3}^{1st}=\sum_{\beta }{{{G}_{3\beta }}}{{V}_{\beta }}=0$, $I_{4}^{1st}=\sum_{\beta }{{{G}_{4\beta }}}{{V}_{\beta }}=0$, given that ${{G}_{31}}={{G}_{32}}$, ${{V}_{1}}=-{{V}_{2}}={V}/{2}$, and $V_3=V_4=0$. However, the second-order currents are nonzero:
\begin{equation}
\begin{split}
I_{3}&=(G_{311} - G_{312}) V^2/2,\\
I_{4}&=(G_{411} - G_{412}) V^2/2.
\end{split} \label{H-cur}
\end{equation}
It is clear from Table \ref{Symmetric relation I} that the second-order current $I_{3}$ is not equal to $I_{4}$ in a system with broken inversion symmetry, and hence the Hall current ${{I}_{H}}=I_{3}-I_{4}\ne 0$. When the applied bias is along the $y$ direction, both the Hall resistance $R_H$ and current $I_H$ are zero in the $x$ direction for the system with mirror symmetry ${{\mathcal{M}}_{x}}$ due to Eq.(\ref{mx}), as discussed in previous cases. In summary, for an inversion-broken system with only the ${{\mathcal{M}}_{x}}$ symmetry described by Eq.(\ref{ham}), nonlinear Hall effect only exists in the $y$ direction and is contributed solely by the second-order nonlinear properties.

The nonlinear Hall transport phenomena from our theoretical analysis via the Landauer-B\"uttiker formula have one-to-one correspondence with the semiclassical Boltzmann approach\cite{L-Fu} at zero frequency. It was argued that~\cite{L-Fu}, for a 2-dimential(2D) TR system with single mirror symmetry, the Berry curvature dipole behaves like a $pseudovector$ in the 2D plane and is forced to be perpendicular to the mirror line. When the driving electric field is aligned with the BCD vector, the flowing currents orthogonal to this driving field are solely contributed by the BCD term and leads to the extrinsic nonlinear Hall effect~\cite{L-Fu}. In vector notation, this second-order Hall current is proportional to $\overrightarrow{D} \cdot \overrightarrow{E}$, with $\overrightarrow{D}$ the BCD vector and $\overrightarrow{E}$ the electric field~\cite{L-Fu}. Therefore, if the applied electric field is perpendicular to the BCD vector, $\overrightarrow{D} \cdot \overrightarrow{E} =0$ and the nonlinear Hall current vanishes. In our situation, for the system determined by Eq.(\ref{ham}), only mirror symmetry $\mathcal{M}_{x}$ is preserved and the mirror line is the $y$-axis. When bias voltages are applied across terminals 1 and 2 that is perpendicular to the mirror line, nonlinear Hall properties from the second-order contributions are observed in the $y$ direction. On the contrary, the Hall effect is absent when applying external voltages in terminals 3 and 4, since the driving field is parallel to the mirror line. Similar to the semiclassical BCD-induced nonlinear Hall effect, we find one-to-one corresponding nonlinear Hall effect in the same 2D massive Dirac Hamiltonian in quantum transport regime, from the symmetry point of view. Since this quantum nonlinear Hall effect is determined by symmetries of the system, we refer it as the symmetry-related nonlinear Hall effect in the following. The one-to-one correspondence between our results and the previous semiclassical study~\cite{L-Fu} confirms that this second-order nonlinear theory based on the nonequilibrium Green's function is appropriate in studying nonlinear Hall effect. Moreover, this quantum transport theory enables us to evaluate the influence of quantum effects such as the internal Coulomb interaction and phase relaxation process, which are presented in the following sections.

\subsection{Numerical results on nonlinear Hall effect}

\begin{figure}[tbp]
\includegraphics[width=9cm, clip=]{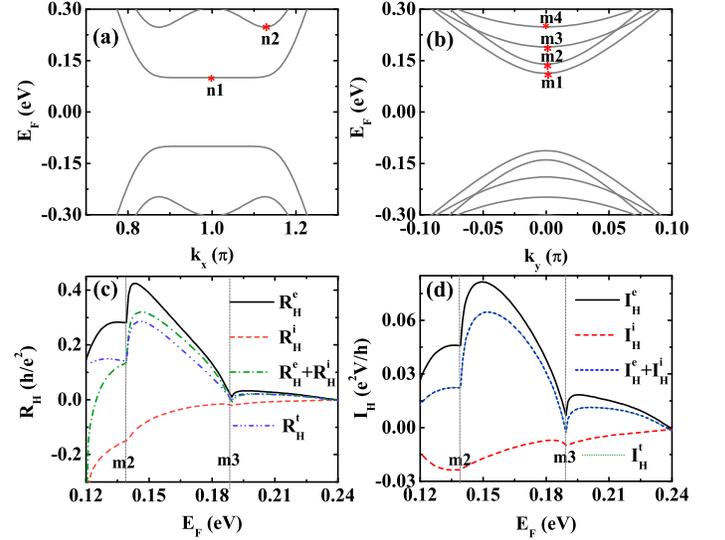}
\caption{\label{Hall resitance}(Color online) (a) and (b): Band structures of the system along $x$ and $y$ directions. The Hall resistance $R_H$ (c) and Hall current $I_H$ (d) versus the Fermi energy $E_F$ in the four-terminal system.}
\end{figure}

In this subsection, we numerically investigate the quantum nonlinear Hall properties of the inversion-broken system described in Eq.(\ref{ham}) on a square lattice. In the calculation, the size of the central region for the four-terminal system shown in Fig.\ref{device} is fixed as $N=L\times L$ with $L = 20$. As a start, band structures of the system along $x$ and $y$ directions are plotted in Figs.\ref{Hall resitance}(a) and \ref{Hall resitance}(b). Since the band structures are symmetric about $E_F=0$, we only study the case of $E_F>0$. To distinguish different subbands, the band edges are labeled with $n\textbf{j}$ and $m\textbf{j}$ for $\textbf{j}=1,2,\cdots $, respectively.

To study the role played by the internal Coulomb potential, we calculate $G_{\alpha \beta \gamma }^{e}$ and $G_{\alpha \beta \gamma }^{i}$ in Eq.(\ref{eq14}) separately, to give the partial Hall resistances $R_{H}^{e}$ and $R_{H}^{i}$, and similarly for the partial Hall currents $I_{H}^{e}$ and $I_{H}^{i}$. $G_{\alpha \beta \gamma }^{e}$ is contributed by the propagating Bloch electrons, and $G_{\alpha \beta \gamma }^{i}$ originates from the induced Coulomb potential. For the total Hall current ${{I}_{H}^{t}}$ calculated from Eq.(\ref{H-cur}) under the closed boundary condition, it is obvious that ${{I}_{H}^{t}}$ is linearly proportional to $G_{\alpha \beta \gamma}$, and $G_{\alpha \beta \gamma} = G_{\alpha \beta \gamma }^{e} + G_{\alpha \beta \gamma }^{i}$ leads to ${{I}_{H}^{t}} = I_{H}^{e} + I_{H}^{i}$. The situation is different for the total Hall resistance ${{R}_{H}^{t}}$. Since the Hall voltage $V_H={V_3} - {V_4}$ is obtained by solving nonlinear equations in Eq.(\ref{H-R}), the sum of $R_{H}^{e}+R_{H}^{i}$ is not equal to ${{R}_{H}^{t}}$.

In calculating $G_{\alpha \beta \gamma }^{i}$, which is contributed by the internal Coulomb potential, the characteristic potential $u_\alpha$ has to be solved. To avoid solving the Poisson equation self-consistently, we use the quasi-neutrality approximation\cite{but1} so that the local charge density is zero, from which the characteristic potential is found to be\cite{but1,sheng},
\begin{equation}
u_\alpha = \frac{dn_\alpha}{dE}/\frac{dn}{dE}. \nonumber
\end{equation}

\begin{figure}[tbp]
\includegraphics[width=8.5cm, clip=]{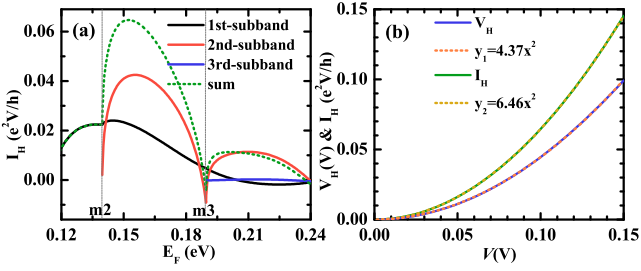}
\caption{\label{Hall resitance-subband}(Color online) (a) Different subband contributions to the Hall current $I_H$ versus the Fermi energy $E_F$. (b) The Hall voltage $V_H$ (at $E_F=0.145$) and Hall current $I_H$ (at $E_F=0.153$) versus the applied bias $V$ along the $x$ direction.}
\end{figure}

Both $I_H$ and $R_H$ can reveal the nonlinear Hall information of the inversion-broken system with TR symmetry. In Figs.\ref{Hall resitance}(c) and \ref{Hall resitance}(d), the Hall resistance and Hall current versus the Fermi energy $E_F$ are presented. The sum of $R_{H}^{e}+R_{H}^{i}$ and $I_{H}^{e}+I_{H}^{i}$ are also plotted for comparison. Several observations are in order. (1). The contribution from the Coulomb interaction is significant and has an opposite sign. It is understandable since the induced Coulomb potential is always against the external bias. (2). For the energy window shown in Fig.\ref{Hall resitance}, the injected electron traversing along the $x$ direction is in the first transmission channel (Fig.\ref{Hall resitance}(a)), while the outgoing electron along the $y$ direction can have several conducting channels ($m1$, $m2$ and $m3$ in Fig.\ref{Hall resitance}(b)) due to the broken $\mathcal{M}_{y}$ symmetry. As the Fermi energy increases, the abrupt change in the Hall resistance or current originates from this subband nature. Moreover, at the third subband threshold $m3$, the total Hall current $I_{H}^{t}$ is negative. If the $I_{H}^{i}$ term contributed from the Coulomb potential is not included, the nonlinear Hall current would remain positive. Hence the induced Coulomb potential leads to a negative $I_{H}^{i}$. (3). The curves of $R_H^{t}$ and $I_{H}^{t}$ in Fig.\ref{Hall resitance}(c) and \ref{Hall resitance}(d) have similar behaviors. $I_{H}^{t}$ increases monotonically in the first subband. As the Fermi energy across the second subband threshold, one more transmission channel is open, giving rise to a jump in the Hall current. Entering the second subband, $I_{H}^{t}$ continues to rise, reaches the maximum and then decreases monotonically until $E_F$ hits the third subband. As shown in Fig.\ref{Hall resitance-subband}(a) that the second subband contribution to $I_H$ is positive. In the third subband, contributions from three transmission channels give a negative Hall current at the third subband threshold (see Fig.\ref{Hall resitance-subband}(a)). As the Fermi energy increases, the first subband contribution to $I_H$ becomes negative while the third subband contribution is two orders of magnitude smaller than that of the first and second subband. Overall, the Hall current behaves similarly as in the second subband. (4). With further increase of $E_F$, the energy bands are getting closer, and the difference between voltages $V_3$ and $V_4$ ( the second-order currents $I_{3}^{2nd}$ and $I_{4}^{2nd}$) caused by the broken mirror symmetry $\mathcal{M}_{y}$ is smaller. As a result, the nonlinear Hall voltage $V_H$ and current $I_H$ approach to zero when the Fermi energy $E_F$ is far away from the gap.

\begin{figure}[tbp]
\includegraphics[width=8.5cm, clip=]{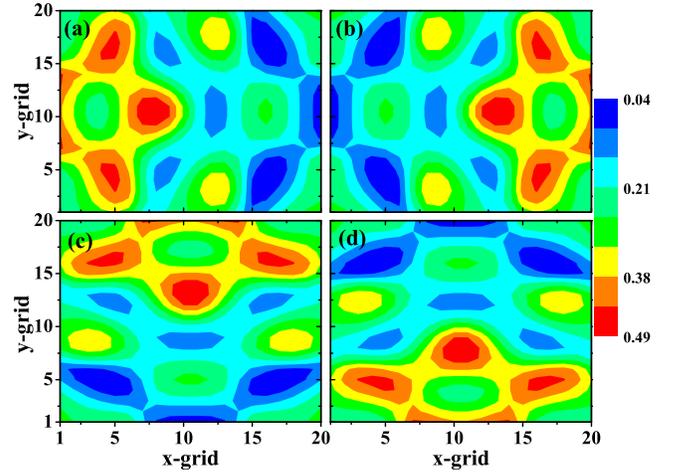}
\caption{\label{char}(Color online) The characteristic potential $u_1$ (a), $u_2$ (b), $u_3$ (c) and $u_4$ (d) when both TR and inversion symmetry are present.}
\end{figure}

The nonlinear characteristics of the Hall properties are shown in Fig.\ref{Hall resitance-subband}(b), where $V_H$ and $I_H$ versus the applied bias voltage along the $x$ direction are plotted. $I_H$ in Fig.\ref{Hall resitance-subband}(b) follows from Eq.(\ref{H-cur}) and increases quadratically with $V$. $V_H$ in Fig.\ref{Hall resitance-subband}(b) is approximately parabolic when $V_1=V/2$ is small. This is because when $V_1$ is small, $V_3$ and $V_4$ are of order $V_1^2$. Therefore one can safely drop quadratic terms of $V_3$ and $V_4$ in Eq.(\ref{H-R}) and immediately obtain $V_3$ and $V_4$ by solving linear equations.

\begin{figure}[tbp]
\includegraphics[width=8.5cm, clip=]{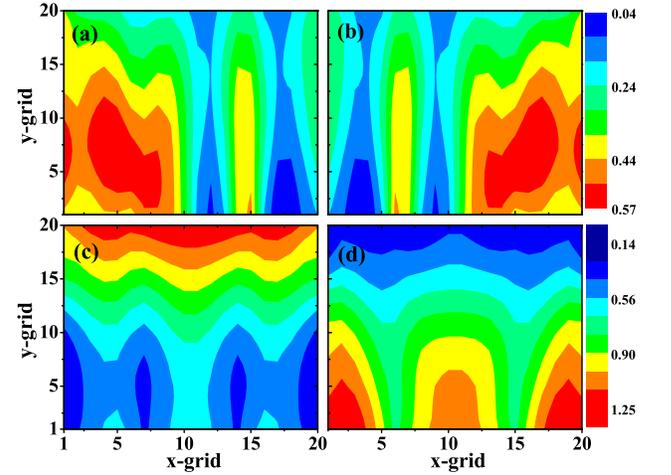}
\caption{\label{char1}(Color online) The characteristic potentials $u_1$ (a), $u_2$ (b), $u_3$ (c) and $u_4$ (d) for the TR system with mirror symmetry $\mathcal{M}_{x}$.}
\end{figure}

To investigate the symmetry of the induced Coulomb potential, we plot in Fig.\ref{char} the characteristic potential $u_\alpha$ ($\alpha=1,2,3,4$) for Case 1 with both TR and inversion symmetries. The characteristic potential $u_\alpha$ is related to the induced nonequilibrium Coulomb potential $U_{neq}$ through $U_{neq}=\sum_{\alpha} {u}_{\alpha} {V}_{\alpha}$ up to the first order in voltage (Eq.(\ref{X2char})). Note that $u_\alpha$ originates from the injection from terminal $\alpha$. As a result, the potential profile of $u_\alpha$ is always higher at the corresponding terminal $\alpha$. From Fig.\ref{char} we see that $u_1$($u_2$) itself has the up-down symmetry, since electrons injected from the left or right terminal experience the mirror symmetry $\mathcal{M}_{y}$ of the system. Clearly, the summation $ u_1 V_1+ u_2 V_2 = (u_1-u_2)V/2$ has both $\mathcal{M}_{x}$ and $\mathcal{M}_{y}$ symmetries, which in fact is the inversion symmetry. Similar observations can be obtained through analyzing the symmetry of $u_3$, $u_4$, and $ u_3 V_3+ u_4 V_4$.

As a result, the induced Coulomb potential, which is the summation of all ${u}_{\alpha} {V}_{\alpha}$, has the inversion symmetry. These facts indicate that the induced Coulomb potential preserves the same symmetry as that of the underlying system. Indeed, this statement is also valid for Case 2, i.e., the system with single mirror symmetry $\mathcal{M}_{x}$. Fig.\ref{char1} depicts the characteristic potential $u_\alpha$ for Case 2. It is clear from $u_3$ and $u_4$ that electrons coming from the up or down terminal experience the mirror symmetry $\mathcal{M}_{x}$. However, $u_1$ and $u_2$ indicate that no spatial symmetry is undergone by electrons injected from the left or right terminal, and $u_1 V_1+ u_2 V_2$ only recovers the $\mathcal{M}_{x}$ symmetry. Hence, when the mirror symmetry $\mathcal{M}_{y}$ is broken, the induced Coulomb potential also has the broken mirror symmetry $\mathcal{M}_{y}$ in real space. This nonequilibrium Coulomb potential in turn gives rise to an extra nonlinear Hall effect (NHE), in addition to the symmetry-related nonlinear Hall effect of this inversion-broken system. The competition between the Coulomb potential induced NHE and the symmetry-related NHE reduce the overall nonlinear Hall response, since the internal Coulomb potential is always against the bias and its contribution to $R_H$ and $I_H$ has opposite sign (Fig.\ref{Hall resitance}).

\subsection{Nonlinear Hall effect in the presence of dephasing}

Next, we study the nonlinear Hall effect in the presence of dephasing. The virtual probe method is used to simulate the dephasing process\cite{but-virtual}. The virtual probe acts as a voltage probe that allows exchange of electrons between the scattering region and the reservoir but forbids the current passing through. The thermalization of electron entering virtual probe by dissipation makes the electron losing its phase memory. Here we show that dephasing can also cause an additional nonlinear Hall effect. In the calculation, we assume that the phase relaxation occurs only in the central region. A voltage probe is attached to each site $\textbf{i}$ with the constraint $I_\textbf{i}=0$ so that dephasing is taken into account from the response of this virtual probe\cite{datta}. The Hamiltonian with virtual probes in momentum space is written as
\begin{equation}
{{H}_{virtual}}=\sum\limits_{\textbf{i},k}{{{\varepsilon }_{k}}}\psi _{\textbf{i}k}^{\dagger }{{\psi }_{\textbf{i}k}}+\sum\limits_{\textbf{i},k}{\left( {{t}_{k}}\psi _{\textbf{i}k}^{\dagger }{{\psi }_{\textbf{i}}}+\textbf{H}.\textbf{c}. \right)}, \nonumber
\end{equation}
where $\psi _{\textbf{i}k}^{\dagger }$ (${{\psi }_{\textbf{i}k}}$) is the creation (annihilation) operator of the electron in the virtual probe, $t_k$ denotes the coupling strength between the virtual probe and the central region. In addition, the retarded self-energy of the virtual probe is $\Sigma _{\textbf{i}}^{r}={-i\Gamma }/{2}\;$ with $\Gamma$ the dephasing strength. The number of virtual probes equals to the size of the central region $N=L\times L$.
Thus, including four real probes, there are total of $N+4$ probes. In the presence of dephasing, we still use open and closed boundary conditions to calculate the Hall resistance and current, respectively. Under the open boundary condition, with $I_3=I_4=0$, $V_1=-V_2={V}/{2}\;$ and $N$ extra boundary conditions $I_\textbf{i}=0$, we can obtain the voltage profile of virtual probes, i.e., the bias ${{\upsilon }_{\textbf{i}}}$ at each virtual probe. Taking the contribution of virtual probes into account, the voltages $V_3$ and $V_4$ are obtained by solving the quadratic equations
\begin{equation}
\begin{split}
{{I}_{3}}&=\sum\limits_{\beta }{{{G}_{3\beta }}}{{V}_{\beta }}+\sum\limits_{\beta \gamma }{{{G}_{3\beta \gamma }}}{{V}_{\beta }}{{V}_{\gamma }}+\sum\limits_{i=1}^{N}{{{G}_{3\textbf{i}}}}{{\upsilon }_{\textbf{i}}}=0,\\
{{I}_{4}}&=\sum\limits_{\beta }{{{G}_{4\beta }}}{{V}_{\beta }}+\sum\limits_{\beta \gamma }{{{G}_{4\beta \gamma }}}{{V}_{\beta }}{{V}_{\gamma }}+\sum\limits_{i=1}^{N}{{{G}_{4\textbf{i}}}}{{\upsilon }_{\textbf{i}}}=0,
\end{split}
\end{equation}
where the conductance from site $\textbf{i}$ to terminal $\alpha $ is given by
\begin{equation}
{{G}_{\alpha i}}=\frac{{{e}^{2}}}{2\pi}Tr\left[ {{\Gamma }_{\alpha}}G^{r}{{\Gamma }_{i}}G^{a} \right]. \nonumber
\end{equation}
\begin{figure}
\includegraphics[width=8.5cm, clip=]{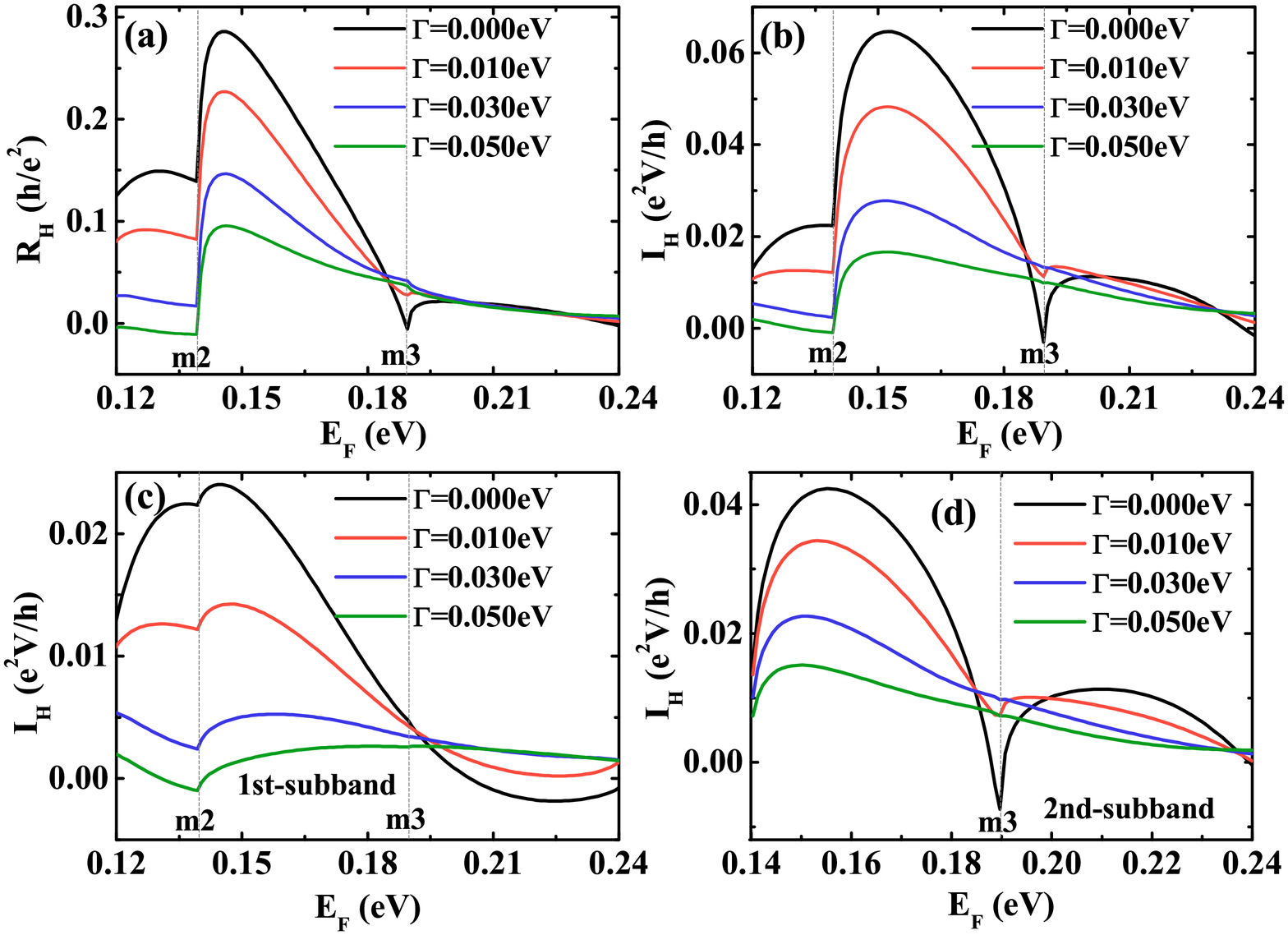}
\caption{\label{dephasing}(Color online) The Hall resistance $R_H$ (a) and Hall current $I_H$ (b) versus the Fermi energy $E_F$ under different dephasing strengths. (c) and (d) show subband contributions to the Hall current.}
\end{figure}

In this way, the Hall resistance ${{R}_{H}}={\left( {{V}_{3}}-{{V}_{4}} \right)}/{{{I}_{1}}}\;$ can be obtained in the presence of dephasing.
For the closed boundary condition, we solve for ${{\upsilon }_{\textbf{i}}}$ at each virtual probe with $V_3=V_4=0$, $V_1=-V_2={V}/{2}\;$ and $N$ extra boundary conditions $I_\textbf{i}=0$.
Once ${{\upsilon }_{\textbf{i}}}$ is calculated, the current $I_3$ and $I_4$ are expressed as
\begin{equation}
\begin{split}
{{I}_{3}}=\sum\limits_{\beta }{{{G}_{3\beta }}}{{V}_{\beta }}+\sum\limits_{\beta \gamma }{{{G}_{3\beta \gamma }}}{{V}_{\beta }}{{V}_{\gamma }}+\sum\limits_{i=1}^{N}{{{G}_{3\textbf{i}}}}{{\upsilon }_{\textbf{i}}},\\
{{I}_{4}}=\sum\limits_{\beta }{{{G}_{4\beta }}}{{V}_{\beta }}+\sum\limits_{\beta \gamma }{{{G}_{4\beta \gamma }}}{{V}_{\beta }}{{V}_{\gamma }}+\sum\limits_{i=1}^{N}{{{G}_{4\textbf{i}}}}{{\upsilon }_{\textbf{i}}},
\end{split}
\end{equation}
from which the Hall current ${{I}_{H}}={{I}_{3}}-{{I}_{4}}$ is obtained under dephasing.

\begin{figure}[tbp]
\includegraphics[width=8.5cm, clip=]{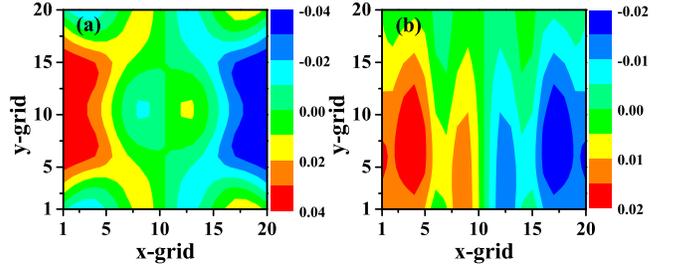}
\caption{\label{voltage}(Color online) Voltage profile of virtual probes for the TR system with inversion symmetry (a) or mirror symmetry $\mathcal{M}_{x}$ (b).}
\end{figure}

Numerical results of the dephasing effect on the nonlinear Hall properties are presented in Fig.\ref{dephasing}. It is known that the dephasing effect can destroy the quantum interference\cite{but-virtual,Y-Wang}. Since the typical quantum interference phenomenon is exhibited in the resonance (resonant peak) and anti-resonance (resonant dip), we expect the phase relaxation process suppresses (enhances) the Hall resistance and current in the resonant region (off-resonant region)\cite{but-pump}. Different subband contributions to $I_H$ are depicted in Fig.\ref{dephasing}(c) and (d). From Fig.\ref{dephasing} we see that the Hall resistance and current decrease as a function of the Fermi energy in the first and second subbands as the dephasing strength increases. For a large enough dephasing strength, $R_H$ and $I_H$ can change sign at the closing of the first subband. Near the second subband, the dip in $R_H$ and $I_H$ are smeared out suggesting the anti-resonance nature of this dip. Furthermore, a moderate strength of dephasing leads to a sign change of $I_H$ at the third subband threshold.

Finally, we show in Fig.\ref{voltage} the voltage profile of virtual probes for Case 1 and 2. Similar to the induced Coulomb potential, the voltage profile also has the same symmetry of the system. Another equivalent description of dephasing is to introduce a complex potential to mimic the inelastic scattering\cite{Y-Wang} or simulate the virtual probe setup. We realize that it is difficult to map the voltage profile to the complex potential. However, in view of the calculated voltage profile of virtual probes, this equivalent complex potential should have the mirror symmetry $\mathcal{M}_{x}$ in real space, which is consistent with the fact that the nonlinear Hall effect is nonzero only along the $y$ direction in the presence of dephasing. Therefore, from this perspective, the complex potential due to dephasing can also lead to the dephasing-induced nonlinear Hall effect. We point out that the phase relaxation process can continuously transform the transport from quantum regime to semiclassical transport, which makes the dephasing-induced nonlinear Hall effect an extrinsic Hall effect.

\subsection{Nonlinear Hall effect at finite temperature}

\begin{figure}[tbp]
\includegraphics[width=8.5cm, clip=]{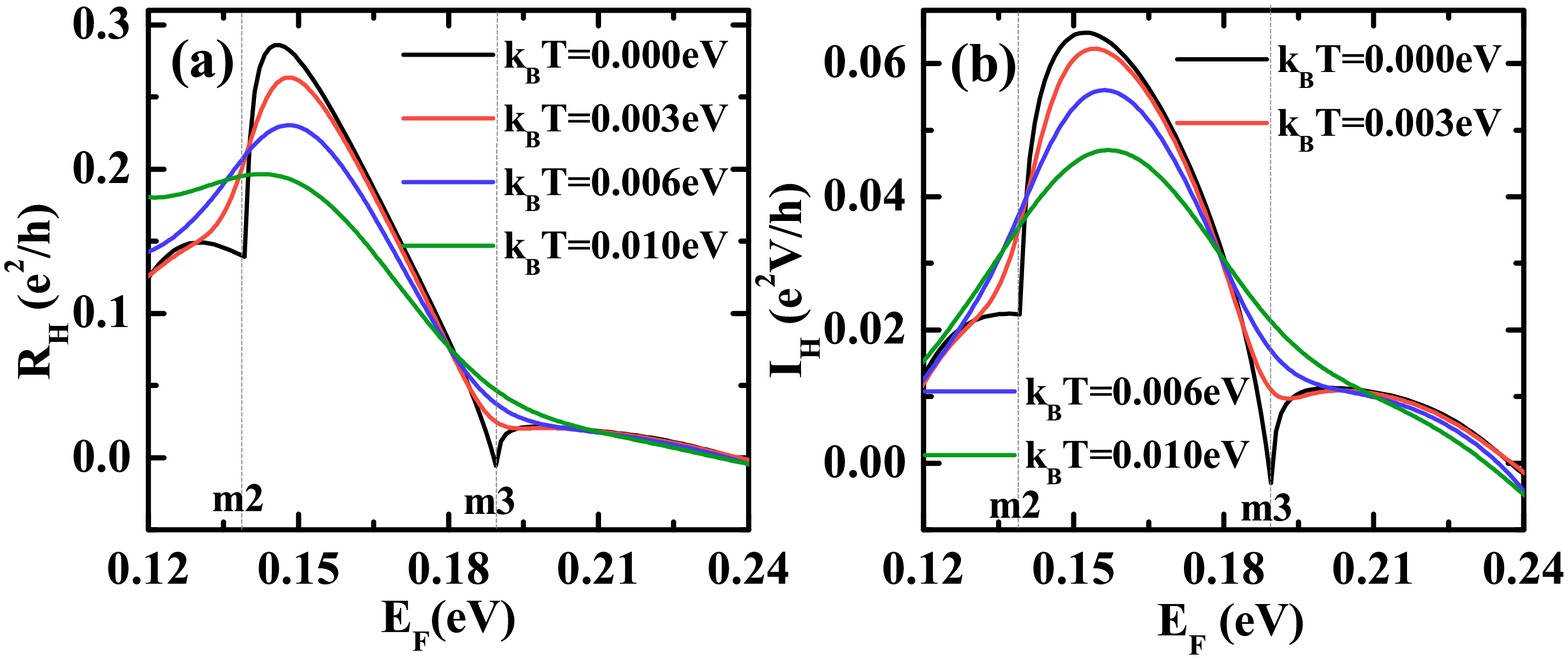}
\caption{\label{non-zero Hall current}(Color online) The Hall resistance $R_H$ (a) and Hall current $I_H$ (b) versus the Fermi energy for different temperatures.}
\end{figure}

In this subsection, we study the nonlinear Hall resistance and Hall current at finite temperature. In Figs. \ref{non-zero Hall current}(a) and \ref{non-zero Hall current}(b), we plot $R_H$ and $I_H$ versus $E_F$ for different temperatures. At finite temperatures, the thermal broadening effect smears out any resonance and anti-resonance. Generally speaking, the Hall resistance and current in the first subband increase with the rising of the temperature while in the second subband they decrease as seen from Fig.\ref{non-zero Hall current}. Around the third subband threshold, $R_H$ and $I_H$ increase with temperature. In addition, the discontinuities at the second and third subband thresholds are also smoothed out. Similar to the dephasing effect, as the temperature is turned on, the negative $I_H$ at the third subband threshold becomes positive.

Since a moving electron also carries energy, the heat current $I^h_\alpha$ is also calculated at finite temperatures. The heat current is defined as the sum of the momentum dependent particle current multiplied by its energy measured from the Fermi level\cite{J-Chen},
\begin{equation}
I^h_{\alpha }= q \int_E (E-E_F -q V_\alpha)\sum_\beta (f_\alpha-f_\beta) T_{\alpha \beta}, \label{heat}
\end{equation}
where $T_{\alpha \beta} = {\rm Tr}[\Gamma_\alpha G^r \Gamma_\beta G^a]$ is the transmission coefficient which depends on the Coulomb potential through Green's functions (see Eq.(\ref{X2gr10})). We emphasize here that the expression Eq.(\ref{heat}) satisfies gauge invariant condition, i.e., the heat current remains the same if the voltage of each terminal is shifted by a constant amount, as discussed in detail in Ref.\onlinecite{J-Chen}. It is easy to see that the total heat current is nonzero,
\begin{eqnarray}
\sum_\alpha I^h_\alpha = \sum_\alpha V_\alpha I_\alpha, \nonumber
\end{eqnarray}
which is just the Joule heating. In particular, for a two-probe system, $\sum_\alpha I^h_\alpha = (V_L-V_R) I_L$.

\begin{figure}[tbp]
\includegraphics[width=8.5cm, clip=]{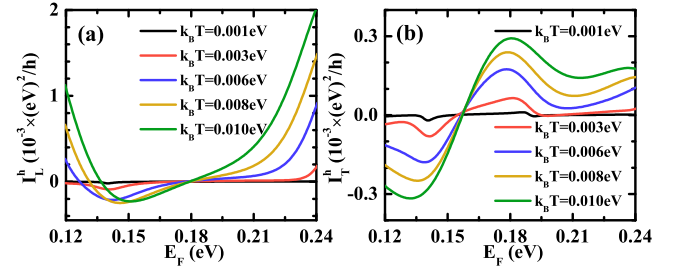}
\caption{\label{heat current}(Color online) The longitudinal heat current ${I_L^h}$ (a) and transverse heat current ${I^h_T}$ (b) versus the Fermi energy for different temperatures.}
\end{figure}

Now, we can fix the voltage at each terminal ${\cal{V}}_\alpha = (V/2,-V/2,0,0)$ as before and expand the heat current in terms of voltage up to the second order, from which we have the following expression for the linear and second-order heat conductances\cite{J-Chen}
\begin{equation}
{{I}^h_{\alpha }}= I_\alpha^{h1} + I_\alpha^{h2}, \nonumber
\end{equation}
where
\begin{eqnarray}
I_\alpha^{h1} =-q\sum_\beta \int_E (E-E_F) (-{\partial_E}f) T_{\alpha \beta } V_\beta, \nonumber
\end{eqnarray}
is the linear heat current and
\begin{eqnarray}
I_\alpha^{h2}&=&q^2 \sum_{\beta\gamma}\int_E (E-E_F) (-{\partial_E}f) {T_{\alpha \beta \gamma }} V_\beta V_\gamma\nonumber \\
&-&(q^2/2) \sum_\beta \int_E (-{\partial_E}f) T_{\alpha \beta } (V_\alpha -V_\beta)^2, \label{heat1}
\end{eqnarray}
is the second-order heat current. Here $T_{\alpha \beta \gamma }$ is defined by $G_{\alpha \beta \gamma } \equiv \int_E (-{\partial_E}f) T_{\alpha \beta \gamma }$ through Eqs.(\ref{eq14}), (\ref{X2g111n}), and (\ref{X2g111n1}). When TR and inversion symmetries are preserved, it is easy to show that $I^h_3-I^h_4=0$ up to the second order in voltage, i.e., no transverse heat current. Similarly, the nonlinear transverse Hall current along $x$ direction is also zero. It suggests that the nonzero transverse heat current is also a measure of the nonlinear Hall effect.

Now we consider the system with mirror symmetry $\mathcal{M}_{x}$, i.e., the Hamiltonian defined in Eq.(\ref{ham}). In this case, when the longitudinal bias voltage is applied, a longitudinal heat current is given by ${I^h_{L}}={I^h_{1}}-{I^h_{2}}$. Moreover, accompanied with the generation of the Hall current, the transverse heat current occurs, which can be expressed as ${I^h_{T}}=I^{h2}_{3}-I^{h2}_{4}$ because the first order heat current in terminals 3 and 4 are zero. In addition, it is easy to show that the second term in Eq.(\ref{heat1}) does not contribute to the transverse heat current due to the symmetry of linear conductance. Fig.\ref{heat current} shows the longitudinal and transverse heat currents versus the Fermi energy $E_F$ for different temperatures. In general, both longitudinal and transverse heat currents increase as the temperature increases and the longitudinal heat current is one order of magnitude larger than the transverse heat current. The transverse heat current in the first subband is negative, while in the second and third subbands it is positive.

\section{Conclusion}

In summary, we have studied the second-order nonlinear Hall resistance and Hall current, as well as the second order longitudinal and transverse heat currents of a four-terminal system with time-reversal symmetry and mirror symmetry $\mathcal{M}_{x}$ in the quantum transport regime. For the same Hamiltonian, we found one-to-one correspondence between these quantum nonlinear Hall properties and the semiclassical BCD-induced nonlinear Hall effect from the symmetry point of view. Quantum effects such as the internal Coulomb interaction and the dephasing mechanism on nonlinear Hall responses have also been investigated. It is found that the nonequilibrium internal Coulomb potential and the voltage profile of virtual probes have the same symmetry as that of the underlying system, which gives rise to two additional nonlinear Hall effects: (1) nonlinear Hall effect induced by internal Coulomb potential and (2) nonlinear Hall effect induced by dephasing effect. These findings are in the quantum transport regime, which are beyond the reach of semiclassical Boltzmann approach.

\section*{acknowledgments}
This work was supported by the National Natural Science Foundation of China (Grant Nos. 12034014 and 12174262), the Natural Science Foundation of Guangdong (Grant No. 2020A1515011418), and the Natural Science Foundation of Shenzhen (Grant Nos. 20200812092737002, JCYJ20190808150409413, and JCYJ20190808115415679).


\begin{thebibliography}{00}
\bibitem{X2foot11}
The injectivity was introduced by M. B\"uttiker in the context of scattering matrix theory and then be expressed with the Green's function\cite{but3}.
\bibitem{X2foot20}
For simplicity, we have neglected terms such as $\partial_E \Sigma^r(E)$.
\end{thebibliography}
\end{document}